\documentstyle[a4,12pt,titlepage]{article}
\newcommand{\x}{\stackrel{\otimes}{,}}

\newcommand{\0}{\nonumber}

\newcommand{\ds}{Drinfel'd-Sokolov }

\newcommand{\3}{{\bf 3}}
\newcommand{\4}{{\bf 3^\ast}}
\newcommand{\mincir}{\raise -2.truept\hbox{\rlap{\hbox{$\sim$}}\raise5.truept
\hbox{$<$}\ }}
\newcommand{\magcir}{\raise -2.truept\hbox{\rlap{\hbox{$\sim$}}\raise5.truept
\hbox{$>$}\ }}
\newcommand{\minmag}{\raise-2.truept\hbox{\rlap{\hbox{$<$}}\raise 6.truept\hbox
{$>$}\ }}
\newcommand{\be}{\begin{equation}}
\newcommand{\ee}{\end{equation}}
\newcommand{\ba}{\begin{eqnarray}}
\newcommand{\ea}{\end{eqnarray}}
\newcommand{\ban}{\begin{eqnarray*}}
\newcommand{\ean}{\end{eqnarray*}}
\newcommand{\brr}{\begin{array}}
\newcommand{\err}{\end{array}}
\newcommand{\bc}{\begin{center}}
\newcommand{\ec}{\end{center}}

\newcommand{\lb}{{\left<\right.}}
\newcommand{\rb}{{\left.\right>}}
\newcommand{\xq}{(q-q^{-1})}
\begin{document}
\begin{titlepage}
\begin{flushright}
SISSA--ISAS 110/92/EP
\end{flushright}
\centerline{\LARGE Quantum $sl_n$ Toda field theories}
\vspace{1.5truecm}
\centerline{\Large L.Bonora,  V.Bonservizi}
\centerline{International School for Advanced Studies (SISSA/ISAS)}
\centerline{Via Beirut 2, 34014 Trieste, Italy}
\centerline{and INFN, Sezione di Trieste.  }
\vspace{5truecm}
\centerline{\large {\bf Abstract}}
\vspace{0.3truecm}
We quantize $sl_n$ Toda field theories in a periodic lattice. We find
the quantum exchange algebra in the diagonal monodromy (Bloch wave) basis
in the case of the defining representation. In the $sl_3$ case we extend
the analysis also to the second fundamental representation. We clarify, in
particular, the relation of Jimbo and Rosso's quantum $R$ matrix with the
quantum $R$ matrix in the Bloch wave basis.
\end{titlepage}

\section{Introduction}

This paper deals with the quantization of Toda field theories based on
a finite Lie algebra in a periodic lattice.

The reason for the interest in Toda field theories is twofold: on the one
hand these theories, which we recall are characterized by a W-algebra
symmetry, underlie a vast set of conformal field theories, in particular
the W minimal models \cite{FL1}, \cite{GeBi}; on the other hand they define
the so-called W-gravities, generalizations of the Liouville theory that
might relate to the most recent results from matrix models and topological
gravity.

While the quantum $sl_2$ Toda theory, i.e. the Liouville theory, has been
analyzed in a number of papers, the attempts to do the same for a general
Toda field theory are fewer and much less complete.

Motivated by the renewed interest in Toda theories, in this paper we want
to analyse in detail the quantization of the $sl_n$, and in particular the
$sl_3$, Toda field theory. The main  difficulty in a continuum context is
that we have to regularize path-ordered exponentials, this being done
usually by introducing a normal ordering. However the two orderings do not
quite match. Here we overcome this difficulty by regularizing the theory
with a lattice cutoff (as suggested by \cite{FTVB}).

The purpose of our paper is to derive the exchange algebra in the Bloch
wave basis, a crucial intermediate step in the calculation of conformal
blocks with diagonal monodromy. We follow the treatment of ref.\cite{BB91}
(see also \cite{B91}). I.e. we start discretizing the chiral and antichiral
\ds linear systems associated with the Toda theory. In this way we
introduce first the exchange algebra in the $\sigma$ basis. The exchange
algebra in this basis is expressed in terms of the quantum $R$ matrix of
Jimbo and Rosso \cite{JR}. At this point  we are very close to the usual
Coulomb gas derivation of conformal blocks (with non-diagonal monodromy).
But instead of proceeding to the calculation of the conformal blocks, we
diagonalize first the monodromy and pass from the $\sigma$ basis to the
$\psi$ basis (Bloch wave basis) through an operator-valued matrix
transformation. This passage guarantees not only periodicity but also
locality in the form explained below. The exchange algebra in the $\psi$
basis is expressed in terms of an $R$ matrix whose entries are in general
functions of zero mode operators. We will show that, up to a diagonal
matrix transformation, this coincides with the $R$ matrix found in
ref.\cite{GR}.

As a bonus of our treatment we are able to answer a question raised in
ref.\cite{CG} (see also \cite{BDF}): what is the relation between the $R$
matrix of Drinfel'd, Jimbo and Rosso and the $R$ matrix that appear in the
Bloch wave basis? The answer is: they coincide up to the transformation
that allows us to pass from the $\sigma$ to the $\psi$ basis.

The paper is organized as follows. In section 1 we review the formulation
of general classical Toda field theories both in the continuum and on the
lattice. In section 3 we first review the recipe for quantizing Toda field
theories on the lattice, found in ref.\cite{BB91}, then we apply it to the
$sl_3$ Toda field theory in the fundamental representations, we find the
quantum exchange algebra in the Bloch wave basis and verify periodicity and
locality. Section 4 is devoted to a comparison of our results with the
exchange algebra obtained previously in ref.\cite{GeBi},\cite{CG} and
\cite{GR}: the results coincide (up to normalization problems). In section
5 we extend the same quantization procedure to the case of the defining
representation of $sl_n$.

\section{Classical Toda Field Theories}

\subsection{In the continuum...}

Let ${\cal G}$ be a simple finite dimensional Lie algebra of rank $n$. We
choose a Cartan subalgebra (CSA) with an orthonormal basis $\{ H_i \}$. The
Toda field equations are
\ba
\partial_{x_+}\partial_{x_-} \Phi \,=\, {\textstyle {1\over 2}}
\sum_{\alpha \;\;simple} e^{2 \alpha (\Phi)} H_\alpha\,,
\label{Toda}
\ea
where $x_\pm = x\pm t$ and the $x$ coordinate lies on a circle; $\Phi$ is
valued in the CSA and $H_\alpha= [E_\alpha, E_{-\alpha}]$, for any simple
root $\alpha$. We showed in \cite{BBT},\cite{ABBP} that the solutions of
(\ref{Toda}) can be obtained from the chirally split \ds linear systems
\cite{DS},
\ba
&&\partial_{x_+} Q_+- (P-{\cal E}_+ )Q_+ \,=\,0\label{DS+} \\[10pt]
&&\partial_{x_-} Q_- +Q_- (\bar{P}-{\cal E}_- )\,=\,0\label{DS-} \,,
\ea
where $P$ and $\bar{P}$ are periodic fields which take
values in the CSA and
\ban
{\cal E}_+ &\!\!=&\!\! \sum_{\alpha ~simple} E_{\alpha}  \\[12pt]
{\cal E}_- &\!\!=&\!\! \sum_{\alpha ~simple} E_{-\alpha} \,.
\ean
$P$ and $\bar P$ have the Poisson brackets
\ba
\{P(x) \x P(y) \}&\!\! =&\!\!- (\partial_x - \partial_y )
\delta(x-y)\, t_0 \label{PP} \\[10pt]
\{P(x) \x \bar{P}(y) \} &\!\!=&\!\!0 \label{PPbar} \\[10pt]
\{\bar{P}(x) \x \bar{P}(y) \} &\!\!=&\!\! (\partial_x - \partial_y)
\delta(x-y) \,t_0\,, \label{PbarPbar}
\ea
where
\ban
t_0\,=\, \sum_i H_i\otimes H_i\,.
\ean
{}From the solution $Q_+(x)$ and $Q_-(x)$ of eqs.(\ref{DS+}) and (\ref{DS-})
normalized by $Q_+(0)=1$, $Q_-(0)=1$, and for any highest weight vector
$|\Lambda^{(r)}\rb$ of ${\cal G}$, we define a basis
$\sigma ,~\bar{\sigma}$
\ba
\sigma^{(r)}(x) &\!\!=&\!\! \lb\Lambda^{(r)}|Q_+(x) \label{sigmar} \\[10pt]
\bar{\sigma}^{(r)}(x) &\!\!=&\!\! Q_-(x) |\Lambda^{(r)}\rb \label{sigmabarr}
\ea
and Bloch wave bases $\psi$, $\bar \psi$
\ba
\psi^{(r)}(x) \,=\,\sigma^{(r)}(x) g \rho, \quad\quad\quad \bar
\psi^{(r)}(x) \,=\,\bar \rho \bar g \bar\sigma^{(r)}(x)\,.
\label{defpsi}
\ea
Here $g$ and $\bar g$ diagonalize the monodromy matrices
\ba
S\,=\,Q_+(2\pi) ,~~~~~~~~~~\bar S\,=\, Q_-(2\pi) \,,
\nonumber
\ea
i.e.
\ba
S&\!\!=&\!\! g\kappa g^{-1},\quad\quad \kappa \,=\, e^{2\pi P_0}
\0\\[10pt]
\bar S &\!\!=&\!\! \bar g^{-1} \bar \kappa \bar g, \quad\quad
\bar \kappa \,=\, e^{-2\pi\bar P_0}\,,
\label{diag}
\ea
$P_0$ and $\bar P_0$ being the zero modes of $P$ and $\bar P$. The matrices
$\rho$ and $\bar\rho$ contain the exponentiated conjugate operators of
$P_0$ and $\bar{P}_0$, respectively. Finally the solutions of the equations
(\ref{Toda}), are reconstructed by means of
\ba
e^{-2\Lambda^{(r)}(\Phi)(x_+,x_-)} \,=\,
\psi^{(r)}(x_+)~ \bar{\psi}^{(r)}(x_-) \,,
\label{solution}
\ea
where $\Lambda^{(r)}$ is the weight corresponding to $|\Lambda^{(r)}\rb$.
It is possible to prove that these solutions are periodic and local,
provided
\ban
\kappa \bar \kappa &\!\!=&\!\! 1 \,,
\ean
which amounts to the condition $P_0 = \bar P_0$.

\subsection{...and on the lattice.}

The formulation of Toda field theories on a (periodic) lattice with
$N$ sites runs parallel to the construction of the above section. We
summarize in the following the recipe found in \cite{BB91}. We limit
ourselves to one chirality, as the other chirality has completely parallel
formulas. Moreover we set $t=0$.

We replace $Q_+(x)$ with a discrete matrix $Q_{n}$ which satisfies the
Poisson bracket
\ba
\{Q_n \x Q_m \}&\!\!=&\!\! Q_n \otimes Q_m\left\{
	\theta(n-m) \left[-r +  Q_m^{-1}\otimes Q_m^{-1}(r-t_0)
	Q_m \otimes Q_m \right]\right.\0\\
	&&+\left. \theta(m-n) \left[ -r +  Q_n^{-1}\otimes Q_n^{-1}(r+
	t_0 )Q_n \otimes Q_n \right] \right\} \,.\label{qnqm}
\ea
If we write
\ban
L_n\,=\,Q_n  Q_{n-1}^{-1}
\ean
and assume
\ban
L_n \,=\, 1+\Delta (P_n -{\cal E}_+ ) + O(\Delta^2) \,,
\ean
where $\Delta$ is the lattice spacing, we obtain eq.(\ref{PP}) in the limit
$\Delta \to 0 $.

Next we define the monodromy matrix $S$ by means of
\ban
Q_{N+n}\,=\,Q_n S \,.
\ean
This must satisfy the following Poisson brackets
\ba
\{Q_n \x S \} &\!\!=&\!\! Q_n\otimes S\, \left(-r +
Q_n^{-1}\otimes Q_n^{-1} \cdot(r+ t_0 ) \cdot Q_n \otimes Q_n
-1\otimes S^{-1}\cdot t_0 \cdot 1\otimes S \right)\label{qnS}   \\[10pt]
\{S \x S \} &\!\!=&\!\! S \otimes S \,\left(-r +
S^{-1}\otimes S^{-1}\cdot r \cdot S \otimes S+ \right. \0 \\
&& \left. \quad\quad\quad + S^{-1}\otimes 1\cdot t_0 \cdot S \otimes
1-1\otimes S^{-1} \cdot t_0 \cdot S\otimes 1 \right)\,.\label{SS}
\ea
In eqs.(\ref{qnqm}), (\ref{qnS}) and (\ref{SS}) the matrix $r$ is
either one of the following classical $r$ matrices
\ban
r^+ &\!\!=&\!\! t_0 + 2 \sum _{\alpha~positive} {{E_\alpha \otimes
E_{-\alpha}}\over {(E_\alpha, E_{-\alpha})}}\\[12pt]
r^- &\!\!=&\!\! -t_0 - 2 \sum _{\alpha~positive} {{E_{-\alpha} \otimes
E_{\alpha}} \over {(E_{-\alpha}, E_{\alpha})}}
\ean

The matrix $\rho$ must satisfy
\ba
\{ Q_n \x \rho\}&\!\!=&\!\! -\alpha Q_n \otimes \rho \cdot t_0
\label{Qrho}\\[10pt]
\{ S\x \rho\}&\!\!=&\!\! -\alpha S\otimes \rho \cdot t_0 - \beta t_0 \cdot S
\otimes\rho\label{Srho}\\[10pt]
\{\rho \x\rho\}&\!\!=&\!\!0\,, \label{rhorho}
\ea
where $\alpha$ and $\beta$ are two arbitrary constants such that $\alpha+\beta
=1$. Since the final results expressed in the Bloch basis do not depend
on the value of these constants, we will choose throughout the  paper
\ban
\alpha\,=\,0
\ean

Finally, given any highest weight vector $|\Lambda^{(r)}\rb$, we can construct
the discrete $\sigma$ basis
\ban
\sigma_n \,=\, \lb\Lambda^{(r)}|Q_n
\ean
and, then, the $\psi$ basis, and verify periodicity and locality of
$\psi_n^{(r)} \bar \psi_n^{(r)}$.

\subsection{The $sl_3$ case}

In the $sl_3$ case we will consider essentially the two fundamental
representations ${\bf 3}$ and ${\bf 3^\ast}$.

The conventions we choose are the following. Let $e_{ij}$ represent
the $3\times 3$ matrix with elements $(e_{ij})_{kl}= \delta_{ik}
\delta_{jl}$. Then, the orthonormal basis for the CSA in the ${\bf 3}$ is
\be
H_1\,=\, \textstyle {1\over {\sqrt 2}}(e_{11}-e_{22}), \quad\quad\quad
H_2 \,=\, \textstyle {1\over{\sqrt 6}} (e_{11}+e_{22}-2e_{33}) \nonumber
\ee
In the ${\bf 3^\ast}$ representation we have
\ban
H_1^\ast \,=\,  \textstyle {1\over {\sqrt 2}}(e_{22}-e_{33})\,,
\quad\quad\quad
H_2^\ast \,=\, \textstyle {1\over{\sqrt 6}} (2e_{11}-e_{22}-e_{33}) \,.
\ean
It is now easy to construct the classical $r$ matrices and see, in
particular, that
\ban
(r^\pm)^{(\bf 3, \bf 3)}\,=\,(r^\pm)^{(\4, \4)}\, ,\quad\quad\quad
(r^\pm)^{(\3, \4)}\,=\, (r^\pm)^{(\4, \3)}\,.
\ean
In general we will append a $^*$ to any object in the $\3$ representation
to denote the corresponding object in the $\4$.

It is perhaps worth writing down the quadratic algebra that can be obtained
from the exchange algebra for the following composed objects
\be
W^{(i)}_n \,=\, \epsilon^{ijkl}\sigma_{n+j}^1\sigma_{n+k}^2
\sigma_{n+l}^3\,,\quad\quad\quad\quad i,j,k,l\,=\,0,1,2,3\,, \label{W}
\ee
where $\epsilon$ is the completely antisymmetric symbol.
In the $\3$ representation we get
\ba
&&\brr{l}
\left\{W_n^{(1)}\, ,\, W_m^{(1)}\right\}\,=\,-\frac{1}{3}W_n^{(1)}W_m^{(1)}
(\delta_{n\,m-3}-\delta_{n\,m-1}+\delta_{n\,m+1}-\delta_{n\,m+3})+
\\[10pt]
{}~~~~~~~~~~~~~~~~~~~~~ +W_n^{(2)}W_{n+2}^{(3)}\delta_{n\, m-1}-
W_{n+1}^{(3)}W_{n-1}^{(2)}\delta_{n\, m+1}
\err\0\\[14pt]
&&\brr{l}
\left\{W_n^{(1)}\, ,\, W_m^{(2)}\right\}\,=\,-\frac{1}{3}W_n^{(1)}W_m^{(2)}
(\delta_{n\,m-3}+\delta_{n\,m-2}-2\delta_{n\,m-1}+
\\[10pt]
{}~~~~~~~~~~~~~~~~~~~~~ +\delta_{n\,m}-\delta_{n\,m+1}+
\delta_{n\,m+2}-\delta_{n\,m+3})+
\\[10pt]
{}~~~~~~~~~~~~~~~~~~~~~ +W_n^{(3)}W_{n+2}^{(3)}\delta_{n\, m-1}-
W_{n+1}^{(3)}W_{n-2}^{(3)}\delta_{n\, m+2}
\err\0\\[14pt]
&&\brr{l}
\left\{W_n^{(1)}\, ,\, W_m^{(3)}\right\}\,=\,-\frac{1}{3}W_n^{(1)}W_m^{(3)}
(\delta_{n\,m-3}+\delta_{n\,m-2}-\delta_{n\,m-1}-\delta_{n\,m+2})
\err\0\\[14pt]
&&\brr{l}
\left\{W_n^{(2)}\, ,\, W_m^{(2)}\right\}\,=\,-\frac{1}{3}W_n^{(2)}W_m^{(2)}
(\delta_{n\,m-3}-\delta_{n\,m-1}+\delta_{n\,m+1}-\delta_{n\,m+3})+
\\[10pt]
{}~~~~~~~~~~~~~~~~~~~~~ -W_n^{(3)}W_{n+1}^{(1)}\delta_{n\, m-1}+
W_n^{(1)}W_{n-1}^{(3)}\delta_{n\, m+1}
\err\0\\[14pt]
&&\brr{l}
\left\{W_n^{(2)}\, ,\, W_m^{(3)}\right\}\,=\,-\frac{1}{3}W_n^{(2)}W_m^{(3)}
(\delta_{n\,m-3}+\delta_{n\,m}-\delta_{n\,m+1}-\delta_{n\,m+2})
\err\0\\[14pt]
&&\brr{l}
\left\{W_n^{(3)}\, ,\, W_m^{(3)}\right\}\,=\,-\frac{1}{3}W_n^{(3)}W_m^{(3)}
(\delta_{n\,m-2}+\delta_{n\,m-1}-\delta_{n\,m+1}-\delta_{n\,m+2})
\err
\label{Wn}
\ea
We remark that if we construct analogous $W_n$'s in the $\psi_n$ basis,
the above algebra does not change.

An analogous quadratic algebra was obtained for the $sl_2$ Toda field
theory on the lattice in refs.\cite{FTVB}. In this case it does not seem to
be possible to construct rational combinations of the $W_n^{(i)}$, which
play the role of discrete generators of the Virasoro and $W_3$ algebra like
the $S_n$ of \cite{FTVB}.

\section{Quantum theory on the lattice.}

\subsection{General formulas}

Here we summarize the recipe for quantization obtained in \cite{BB91}.
For any matrix $X$ we use the notation
\ban
X_{1}\,=\,X\otimes I  \,, \quad\quad\quad
X_{2}\,=\, I \otimes X \,.
\ean

The quantum analogue of $Q_n$ must satisfy:

\ba
R_{12} Q_{1n} Q_{2n} &\!\!=&\!\! Q_{2n} Q_{1n} R_{12}\,,
\label{QnQnq}\\[10pt]
Q_{1n}Q_{1m}^{-1}A_{21}R_{12}Q_{1m}Q_{2m} &\!\!=&\!\! Q_{2m}Q_{1n}R_{12}\,,
\quad\quad\quad n>m\,,
\label{QQq}
\ea
with $n,m<N$ and
\ban
A_{12}&\!\!=&\!\! e^{\textstyle{i\hbar}t_0 }
\\[10pt]
R_{12}&\!\!=&\!\! 1-{\textstyle{i\hbar}}r_{12} +O(\hbar^2 )
\ean
$R_{12}$ must satisfy the quantum Yang-Baxter equation. Hereafter we denote
$R_{12}^{\pm}$ the two solutions of the Yang-Baxter equation whose
classical limits are $r_{12}^\pm$, respectively. We have
$R_{12}^-=(R_{12}^+)^{-1}$.


In a periodic lattice with $N$ sites we introduce the quantum monodromy
matrix via
\ban
Q_{n+N}\,=\,Q_nBS\,,
\ean
where
\ban
B\,=\,e^{{{i\hbar}\over 2}\sum_iH_i^2}\,.
\ean
Here, $S$ must satisfy
\ba
R_{12} S_1 A_{12} S_2 &\!\!=&\!\! S_2 A_{12} S_1 R_{12}
\label{RSAS}\\[10pt]
A_{12} R_{12} Q_{1n} Q_{2n} &\!\!=&\!\! Q_{1n}S_1^{-1} Q_{2n} A_{12} S_1
R_{12}\,. \label{QASR}
\ea
Finally, the quantum matrix $\rho \in \exp ({\cal H})$ has the
following properties
\ba
\rho_1 \rho_2 &\!\!=&\!\! \rho_2 \rho_1 \label{qrhorho} \\[10pt]
A_{12}S_1 \rho_2 A_{12} &\!\!=&\!\! \rho_2 S_1 \label{qSrho} \\[10pt]
Q_{1n} \rho_2 &\!\!=&\!\! \rho_2 Q_{1n}\,. \label{qQrho}
\ea

In the above equation we understood the representation labels. They can
be supplied in an obvious way.

Let us define now the quantum $\sigma$ basis
\be
\sigma_n^{(r)} \,=\, \lb\Lambda^{(r)}| Q_n \label{qsigma}\,.
\ee
{}From eq.(\ref{QQq}) we obtain the exchange algebra
\ba
&&\sigma_{1n}^{(r)} \sigma_{2m}^{(r')}\,=\,\sigma_{2m}^{(r')}
\sigma_{1n}^{(r)} (R_{12}^+)^{(r,r')}, \quad\quad\quad n>m \0\\[10pt]
&&\sigma_{1n}^{(r)} \sigma_{2m}^{(r')}\,=\,\sigma_{2m}^{(r')}
\sigma_{1n}^{(r)} (R_{12}^-)^{(r,r')}, \quad\quad\quad n<m \,.
\label{qsigmasigma}
\ea
{}From eq.(\ref{QnQnq}) we can also obtain the exchange algebra for $n=m$.

\subsection{Quantum $sl_3$ Toda field theory on the lattice}

\subsubsection{The quantum exchange algebra}

The purpose of this subsection is to use the general formulas of the
previous subsection to define the Block wave basis $\psi_n^{(r)}$ and find
the relevant exchange algebra in the $sl_3$ case.

We start with the quantum $R$ matrix to be inserted in the above formulas.
{}From the universal $R$ matrix for $sl_n$ as given in \cite{JR} we obtain
\be
(R_{12}^+)^{(\3,\3)}_{ij,kl}\,=\,(R_{12}^+)^{(\4,\4)}_{ij,kl}\,
=\,\left\{
\brr{cc} q^{{2\over 3}}\quad\quad\quad &\scriptstyle i=j=k=l\\
         q^{-{1\over 3}}\quad\quad\quad &\scriptstyle i=k\neq j=l\\
         q^{-{1\over 3}}x\quad\quad &\scriptstyle i=l<j=k\\
         0\quad\quad\quad &\mbox{\small\rm otherwise}    \err
\right.\label{R}
\ee
and
\be
(R_{12}^+)^{(\3, \4)}_{ij,kl}\,=\,(R_{12}^+)^{(\4, \3)}_{ij,kl}\,=\,
\,\left\{
\brr{cc} q^{{1\over 3}}\quad\quad\quad &\scriptstyle i=j,~~k=l,~~i+j\neq 3\\
         q^{-{2\over 3}}\quad\quad\quad &\scriptstyle i=k,~~ j=l,~~i+j=3\\
         q^{-{2\over 3}}x\quad\quad &\scriptstyle i=l=1,~j=3,~k=2~~
         {\rm and}~~i=j=2,~k=3,~l=1\\
         -q^{-{2\over 3}}x\quad\quad &\scriptstyle i=l=1,~j=k=3\\
         0\quad\quad\quad &\mbox{\small\rm otherwise}    \err
\right. \label{R*}
\ee
where $q= e^{i\hbar}$, while, here and hereafter, we denote $x=q-q^{-1}$.
Using these equations we can easily derive the exchange algebra
in the $\sigma$ basis, which will not be written down explicitly here.

As a first step in the direction of the $\psi$ basis, we have to diagonalize
the upper triangular monodromy matrices $S$ and $S^*$ in the two fundamental
representations
\ban
S\,=\,\left( \brr{ccc} A & D & F\\
                0 & B & E\\
                0 & 0 & C\err\right),\quad\quad\quad
S^*\,=\,\left( \brr{ccc} A^* & D^* & F^*\\
                0 & B^* & E^*\\
                0 & 0 & C^*\err\right)\,.
\ean
Eqs.(\ref{RSAS}) and (\ref{QASR})  allow us to compute the commutation
relations
of the entries of $S$ and $S^*$ among themselves and with the component of
the $\sigma_n$'s. One finds that the diagonal elements $A$ $B$ $C$ and
$A^*$ $B^*$ $C^*$ commute with everything (except $\rho$, see below)
while, for example,
\be
DF\,=\,q^{-1}\,FD \quad\quad\quad EF\,=\,q\,FE
\quad\quad\quad ED\,=\,q^{-1} DE+x\,BF
\label{eq:smsm}
\ee
and
\ban
\brr{rlrl}
D\sigma^1_n &\!\!=\, q^{-1}\,\sigma_n^1D+x\,\sigma^2_nA
\quad\quad\quad &
D\sigma_n^2 &\!\!=\, q\,\sigma_n^2D  \\[10pt]
E\sigma_n^1 &\!\!=\, \sigma_n^1 & E\sigma^2_n &\!\!=\, q^{-1}\,
\sigma_n^2 E+x\,\sigma_n^3 B \\[10pt]
F\sigma_n^1 &\!\!=\, q^{-1}\,\sigma_n^1 F+x\,\sigma^3_n A &
F\sigma_n^2 &\!\!=\, \sigma_n^2 F+x\,\sigma_n^3 D
\err
\ean
\ba
D\sigma^3_n &\!\!=\!\!& \sigma_n^3D
\0\\[10pt]
E\sigma_n^3 &\!\!=\!\!& q\,\sigma_n^3E
\0\\[10pt]
F\sigma_n^3 &\!\!=\!\!& q\,\sigma_n^3F\,.
\label{eq:sigsm}
\ea
We will not write down here the remaining relations, except for
\be
AC^*\,=\, A^*C\,=\,BB^*\,.\label{constr1}
\ee
This relation explains why we can express the exchange algebras in terms
of zero modes $A,B,C$ only (see Appendix).

Next we diagonalize $S$ and $S^*$ with upper triangular matrices $g$ and
$g^*$, respectively, which have unit entries in the main diagonals. That is
to say
\ban
S\,=\, g\kappa g^{-1},\quad\quad\quad S^*\,=\, g^* \kappa^* (g^*)^{-1}\,,
\ean
$\kappa$ and $\kappa^*$ being diagonal matrices, whose main diagonals
coincide with the main diagonal of $S$ and $S^*$, i.e. $A,B,C$ and
$A^*,B^*, C^*$, respectively. It is immediate to compute the commutators of
the entries of $g$ and $g^*$ with all the operators introduced so far.

Next we introduce the matrix $\rho$ and the analogous $\rho^*$.

Analogously for the conjugate variables to the zero modes
\be
\rho\,=\,\left(
\brr{ccc}          \rho_1 &&                  \\
                          & \rho_2            \\
                          &        & \rho_3   \err
\right),
\quad\quad\quad
\rho^*\,=\,\left(
\brr{ccc}          \rho^{*}_1 &&                  \\
                          & \rho^{*}_2            \\
                          &        & \rho^{*}_3   \err\,.
\right)
\label{rho}
\ee

Due to eqs.(\ref{qrhorho},\ref{qQrho}) and (\ref{qSrho}), $\rho_i$ and
$\rho_i^*$ with $i=1,2,3$ commute among themselves and with all the operators
we introduced so far, except for the elements of $S$ and $S^*$. For these
we have
\be
S_{ij}\rho_k \,=\, \rho_k S_{ij} q^{{6\delta_{i,k} -2}\over 3}
\quad\quad\quad
S_{ij}\rho^*_k \,=\, \rho^*_k S_{ij} q^{{2-6\delta_{i+k,4} }\over 3}
\label{Sijrho}
\ee
and two more equations which can be obtained by ``starring" these two (with
the understanding that this formal $^*$ operation is involutive).

After introducing the complete set of operators of the theory, we can draw
some immediate useful conclusions. First, $ABC$ and $A^*B^*C^*$ belong to
the center of the theory. Second, $\rho_2\rho_2^*$, $\rho_1\rho_3^*$ and
$\rho_3\rho_1^*$ also commute with everything else. Therefore, we can and
will henceforth impose
\be
ABC\,=\,1\,=\,A^*B^*C^*\label{const2}
\ee
and
\be
\rho_2\rho_2^*\,=\,\rho_1\rho_3^*\,=\, \rho_3\rho_1^*\,.\label{const3}
\ee
This will allow us to simplify many formulas. In particular, (\ref{const2})
allows us to parametrize the zero modes as follows:
\be
\brr{ll}
A\,=\,q^{\frac{N-1}{3}}e^{2\pi (\frac{1}{\sqrt{2}}p_0^1+
        \frac{1}{\sqrt{6}}p_0^2) }\,,\quad\quad\quad &
A^*\,=\,q^{\frac{N-1}{3}}e^{2\pi (\sqrt{\frac{2}{3}}p_0^2)}\,,
\\[12pt]
B\,=\,q^{\frac{N-1}{3}}e^{2\pi (-\frac{1}{\sqrt{2}}p_0^1+
        \frac{1}{\sqrt{6}}p_0^2) }\,, &
B^*\,=\,q^{\frac{N-1}{3}}e^{2\pi (\frac{1}{\sqrt{2}}p_0^1-
        \frac{1}{\sqrt{6}}p_0^2)} \,,
\\[12pt]
C\,=\,q^{\frac{N-1}{3}}e^{2\pi (-\sqrt{\frac{2}{3}}p_0^2)}\,, &
C^*\,=\,q^{\frac{N-1}{3}}e^{2\pi (-\frac{1}{\sqrt{2}}p_0^1-
        \frac{1}{\sqrt{6}}p_0^2)}\,,
\err
\label{zm}
\ee
in agreement with the conventions chosen for the CSA basis and with our
quantization procedure.

Now what remains to be done is to define the quantum Block wave basis
\be
\psi_n\,=\, \sigma_n g \rho,\quad\quad\quad \psi_n^*\,=\,
\sigma_n^* g^*\rho^*
\label{psi}
\ee
and compute its exchange algebra. The calculation is long but uneventful
and the result has the form
\ba
\psi_{1n}\psi_{2m}\,=\, \psi_{2m}\psi_{1n}(R_{12}^\pm(p_0))^{(\3,\3)},
\quad\quad\quad
\left\{\brr{cc} + & n>m \\
                - & n<m \label{50}\err\right.
\ea
\ba
\psi_{1n}\psi^*_{2m}\,=\, \psi^*_{2m}\psi_{1n}(R_{12}^\pm(p_0))^{(\3,\4)},
\quad\quad\quad
\left\{\brr{cc} + & n>m \\
                - & n<m \label{51} \err\right.\,,
\ea
where the argument $p_0$ is to remember the dependence on the zero modes.
The entries of the (zero mode dependent) quantum $R$ matrix\footnote{Here
we enlarge the notion of $R$ matrix. Indeed the exchange matrices of
eqs.(\ref{50}) and (\ref{51}) are not solutions of the YBE's, but of a
modified version of them.} in the Bloch wave basis are written down
explicitly in Appendix A.1 and A.2.

This completes our proof about the relation between the quantum $R$ matrix
of Jimbo and Rosso and the quantum $R$ matrix in the Bloch wave basis. Such
a relation cannot be envisaged as an (operator-valued) similarity
transformation since, for example in (\ref{50}), $g_1\rho_1$ does not
commute with $\psi_{2m}$. We can only say that the relation is specified by
the operator-valued change of basis (\ref{psi}).

In order to discuss periodicity and locality, one has to repeat everything
for the discretization of the antichiral half, in order to calculate the
exchange algebra of
\be
\bar\psi_n\,=\,\bar\rho\bar g \bar\sigma_n  ,
\quad\quad\quad \bar\psi_n^*\,=\,\bar \rho^*\bar g^* \bar \sigma_n^*
\label{barpsi}
\ee
The result of this calculation can also be found in Appendix A.1 and A.2.

\subsubsection{Periodicity and locality.}

In analogy with the continuum case, we define
\ban
e^{-\varphi_n}\,=\, \psi_n \bar \psi_n, \quad\quad\quad
e^{-\varphi^*_n}\,=\, \psi^*_n \bar \psi^*_n\,.
\ean
We find
\ban
&&e^{-\varphi_{n+N}}\,=\, q^{4\over 3} \left( A\bar A \psi_n^1 \bar \psi_n^1+
B\bar B \psi_n^2 \bar \psi_n^2+ C\bar C \psi_n^3 \bar \psi_n^3 \right)
\\[10pt]
&&e^{-\varphi^*_{n+N}}\,=\, q^{4\over 3} \left( A^*\bar A^* \psi_n^{*1} \bar
\psi_n^{*1}+ B^*\bar B^* \psi_n^{*2} \bar \psi_n^{*2}+ C^*\bar C^* \psi_n^{*3}
\bar  \psi_n^{*3} \right)\,.
\ean
Since $A\bar A, B\bar B, C\bar C, A^*\bar A^*, B^*\bar B^*, C^*\bar C^*$
commute with all the operators of the theory, we can project out of the
full Hilbert space ${\cal H}$ the subspace ${\cal H}_0$ where
\be
A\bar A\,=\, B\bar B\,=\, C\bar C\,=\, A^*\bar A^*\,=\, B^*\bar B^*\,=\,
C^*\bar C^*\,=\, q^{-{4\over 3}}\,.\label{constraint}
\ee
In ${\cal H}_0$ both $e^{-\varphi_n}$ and $e^{-\varphi^*_n}$ are
periodic.

To prove locality, we compute
\ban
\relax [e^{-\varphi_n}, e^{-\varphi_m}]&\!\!=&\!\! x \,
\frac {B\bar B -A\bar A} {(\bar B-\bar A)(B-A)}
\left( \psi^1_n \psi^2_m \bar \psi^2_n \bar \psi^1_m -
                  \psi^1_m \psi^2_n \bar\psi^2_m \bar\psi^1_n\right)+\\
&\!\!+&\!\! x\,\frac {C\bar C -A\bar A} {(\bar C-
\bar A)(C-A)} \left( \psi^1_n \psi^3_m \bar \psi^3_n \bar \psi^1_m -
                  \psi^1_m \psi^3_n \bar\psi^3_m \bar\psi^1_n\right)+\\
&\!\!+&\!\! x\,\frac {C\bar C -B\bar B} {(\bar C-
\bar B)(C-B)} \left( \psi^2_n \psi^3_m \bar \psi^3_n \bar \psi^2_m -
                      \psi^2_m \psi^3_n \bar\psi^3_m \bar\psi^2_n\right)
\ean
Therefore the commutator vanishes in the subspace ${\cal H}_0$. The same
conclusion holds if we consider $[e^{-\varphi_n^*}, e^{-\varphi_m^*}]$.
Next let us consider $[e^{-\varphi_n}, e^{-\varphi_m^*}]$. This commutator
is more complicated than the previous ones. However it can be proven that
it is a combination of terms each of which factorize out either $B\bar B -
A\bar A$ or $C\bar C - A\bar A$ or $C\bar C - B\bar B$. Therefore we
again can conclude that, in ${\cal H}_0$,
\ban
[e^{-\varphi_n}, e^{-\varphi_m^*}]\,=\,0
\ean

This completes the derivation of our result as far as the $sl_3$ Toda field
theory is concerned. It is perhaps useful spending a few words to give the
reader the coordinates of this result in the prospect of evaluating
correlation functions. The lattice analogues of the conformal blocks are
given by expressions like (we consider here only the $\3$ representation)
\ban
\lb\theta_\infty|\psi_{n_1}\psi_{n_2}\cdots \psi_{n_k}|\theta_0\rb
\ean
and
\ban
\lb\bar\theta_\infty|\bar\psi_{n_1}\bar\psi_{n_2}\cdots \bar\psi_{n_k}
|\bar\theta_0\rb
\ean
where the $\theta$ states tend to the corresponding conformal vacua in the
continuum limit. Putting together the two halves, we can compute
\be
\lb e^{-\varphi(x_1)}e^{-\varphi(x_2)}\ldots e^{-\varphi(x_k)}\rb\,,
\label{cf}
\ee
where $e^{-\varphi(x)}$ is the continuum limit of $e^{-\varphi_n}$.
Single-valuedness and locality of (\ref{cf}) is then guaranteed by the
condition (\ref{constraint}).

\section{Comparison with previous results.}

In the previous section we computed the exchange algebra for the $sl_3$
Toda field theory in a periodic lattice. Since this algebra does not depend
on the lattice spacing, we can immediately translate it into a continuous
language by the simple replacements
\ban
\psi^i_n \to \psi^i(x)\,, \quad\quad \psi^{*i}_n \to \psi^{*i}(x)\,
,\quad\quad {\rm etc.}, \quad\quad \theta(n-m)\to \theta(x-y)\,.
\ean
In this way we can compare our results with those of ref.\cite{GeBi} (see
also \cite{CG} and, for the specific case of $sl_3$, \cite{GR}). There,
following a different approach, the $sl_{l+1}$ exchange algebra in the
Bloch wave basis for the defining representation was calculated to be
\be
\brr{l}
\phi_j(\sigma)\phi_j(\sigma^{'}) \,=\, e^{-i\hbar\epsilon\,l/l+1}\,
\phi_j(\sigma^{'})\phi_j(\sigma)\,,
\\[12pt]
\phi_j(\sigma)\phi_k(\sigma^{'}) \,=\, e^{i\hbar\epsilon/l+1}
{sin(\hbar(\varpi_{jk}+1)\over sin(\hbar\varpi_{jk})}\,
\phi_k(\sigma^{'})\phi_j(\sigma)+
\\[10pt]\quad\quad\quad\quad\quad\quad
+e^{i\hbar\epsilon/l+1}{sin(\hbar)e^{-i\hbar\varpi_{jk}}\over
sin(\hbar\varpi_{ij})}\,\phi_j(\sigma^{'})\phi_k(\sigma)\,,
\err
\quad\quad\quad\mbox{\small$\epsilon\,=\,sign(\sigma-\sigma^{'})$}
\label{gervais}
\ee
where
\ban
\varpi_{jk}\,=\,(\lambda_k-\lambda_j)\cdot\vec{\varpi} \quad\quad\quad
\lambda_i&\!\!\!=&\!\!\!\mbox{\small\rm weights of the defining
                              representation}
\\
\vec{\varpi}&\!\!\!=&\!\!\!-\frac{i}{2}\sqrt{\frac{2\pi}{\hbar}}\,\Lambda_0~~,
           ~~~~\Lambda_0\,\!=\,\!\sum_{i=1}^n\Lambda_i\tilde{p}_0^i
\\
\Lambda_i&\!\!\!=&\!\!\!\mbox{\small\rm fundamental weights}
\ean
Here, the zero modes $\tilde{p}_0^i$ ($i=1,\dots,l$) correspond to the
weight space basis consisting of the simple root system
$\{\alpha_i\}_{i=1,\dots,l}$ (that is to say, to the basis
$\{h_i\}_{i=1,\dots,l}$ of the CSA, $h_i=e_{ii}-e_{i+1\,i+1}$). Considering
as an example the case of $sl_3$, the $\tilde{p}_0^{1,2}$ are related to
the zero modes introduced in eqs.(\ref{zm}) by the linear
transformation
\be
\tilde{p}_0^1\,=\,\sqrt{\frac{4\pi}{\hbar}}(\sqrt{2}p_0^1) \quad\quad\quad
\tilde{p}_0^2\,=\,\sqrt{\frac{4\pi}{\hbar}}(-\frac{1}{\sqrt{2}}p_0^1+
 \sqrt{\frac{3}{2}}p_0^2)\,.
\label{eq:p0rel}
\ee
Taking into account such a rotation, together with the usual identification
$q=e^{-i\hbar}$, we obtain the relations
\be
\brr{ll}
\xq{A\over B-A}\,=\,-{sin(\hbar) \over sin(\hbar\varpi_{12})}
       e^{-i\hbar\varpi_{12}}\,, \quad\quad
&
\xq{B\over B-A}\,=\,{sin(\hbar)\over sin(\hbar\varpi_{21})}
       e^{-i\hbar\varpi_{21}}\,,
\\[12pt]
\xq{B\over C-B}\,=\,-{sin(\hbar)\over sin(\hbar\varpi_{23})}
       e^{-i\hbar\varpi_{23}}\,,
&
\xq{C\over C-B}\,=\,{sin(\hbar)\over sin(\hbar\varpi_{32})}
       e^{-i\hbar\varpi_{32}}\,,
\\[12pt]
\xq{A\over C-A}\,=\,-{sin(\hbar)\over sin(\hbar\varpi_{13})}
       e^{-i\hbar\varpi_{13}}\,,
&
\xq{C\over C-A}\,=\,{sin(\hbar)\over sin(\hbar\varpi_{31})}
       e^{-i\hbar\varpi_{31}}\,.
\err
\ee
This allows us to identify the off-diagonal coefficients of the operator
algebra (\ref{gervais}) with the corresponding elements of the zero mode
dependent $R$ matrices of eq.(\ref{50}) (see Appendix A.1). As for the
diagonal entries, it is important to remark that they coincide only up to a
change in the normalization of the Bloch wave basis. Indeed, in order to
reproduce the operator algebra of Appendix A.1, the vertex fields $\phi_i$,
$i=1,2,3$, should be multiplied by the factors
\be
\brr{l}
c_1(\tilde{p}_0)\,=\,\left[sin(\hbar\varpi_{13})\right]^a
\left[sin(\hbar\varpi_{12})\right]^{1-a}\,,
\\[12pt]
c_2(\tilde{p}_0)\,=\,\left[sin(\hbar(\varpi_{12}+1))\right]^a
\left[sin(\hbar\varpi_{23})\right]^{1-a}\,,
\\[12pt]
c_3(\tilde{p}_0)\,=\,\left[sin(\hbar(\varpi_{23}+1)\right]^a
\left[sin(\hbar(\varpi_{13}+1))\right]^{1-a}\,,
\err
\label{eq:norm}
\ee
respectively, where $a$ is an arbitrary parameter. Since the $c_i$'s depend
only on the zero modes, this operation does not modify the monodromy
behaviour of the fields, which still constitute a Bloch wave basis. After
this transformation, while the off-diagonal elements remain unchanged, the
diagonal ones take the expressions\footnote{Following ref.\cite{GR}, we
use for the $R$ matrix elements the notation
$\phi_i\phi_j=\sum_{kl}R_{ij}^{kl}\phi_k\phi_l$.}
\ban
&&\brr{l}
R_{12}^{21}(\tilde{p}_0)\,\to\,{sin(\hbar\varpi_{12})\over
          sin[\hbar(\varpi_{12}+1)]}\,R_{12}^{21}(\tilde{p}_0)\,=\,
          e^{i\hbar\epsilon/3}\,=\,q^{-\frac{1}{3}}\,,
\err\\[14pt]
&&\brr{l}
R_{23}^{32}(\tilde{p}_0)\,\to\,{sin(\hbar\varpi_{23})\over
          sin[\hbar(\varpi_{23}+1)]}\,R_{23}^{32}(\tilde{p}_0)\,=\,
          e^{i\hbar\epsilon/3}\,=\,q^{-\frac {1}{3}}\,,
\err\\[14pt]
&&\brr{l}
R_{13}^{31}(\tilde{p}_0)\,\to\,{sin(\hbar\varpi_{13})\over
          sin[\hbar(\varpi_{13}+1)]}\,R_{13}^{31}(\tilde{p}_0)\,=\,
          e^{i\hbar\epsilon/3}\,=\,q^{-\frac {1}{3}}\,,
\err\\[14pt]
&&\brr{l}
R_{21}^{12}(\tilde{p}_0)\,\to\,{sin[\hbar(\varphi_{12}+1)]\over
          sin(\hbar\varpi_{12})}\,R_{21}^{12}(\tilde{p}_0)\,=\,
  \\[10pt]
  \quad\quad=\,e^{i\hbar\epsilon/3}\left[cos^2(\hbar)-sin^2(\hbar)
         {cos^2(\hbar\varpi_{12})\over sin^2(\hbar\varpi_{12})}\right]\,=\,
         q^{-\frac{1}{3}}\,\left[1-\xq^2{AB\over (B-A)^2}\right]\,,
\err\\[14pt]
&&\brr{l}
R_{32}^{23}(\tilde{p}_0)\,\to\,{sin[\hbar(\varphi_{23}+1)]\over
          sin(\hbar\varpi_{23})}\,R_{32}^{23}(\tilde{p}_0)\,=\,
  \\[10pt]
  \quad\quad=\,e^{i\hbar\epsilon/3}\left[cos^2(\hbar)-sin^2(\hbar)
         {cos^2(\hbar\varpi_{23})\over sin^2(\hbar\varpi_{23})}\right]\,=\,
         q^{-\frac{1}{3}}\,\left[1-\xq^2{BC\over (C-B)^2}\right]\,,
\err\\[14pt]
&&\brr{l}
R_{31}^{13}(\tilde{p}_0)\,\to\,{sin[\hbar(\varphi_{13}+1)]\over
          sin(\hbar\varpi_{13})}\,R_{31}^{13}(\tilde{p}_0)\,=\,
  \\[10pt]
  \quad\quad=\,e^{i\hbar\epsilon/3}\left[cos^2(\hbar)-sin^2(\hbar)
         {cos^2(\hbar\varpi_{13})\over sin^2(\hbar\varpi_{13})}\right]\,=\,
         q^{-\frac{1}{3}}\,\left[1-\xq^2{AC\over (C-A)^2}\right]\,,
\err
\ean
while the entries $R_{ii}^{ii}$ do not transform.

The same can be repeated when, instead of the representations $\3$ and $\3$,
we have $\4$ and $\4$ or $\3$ and $\4$.

In conclusion, we have shown that the $R$ matrices in the Bloch wave basis
of \cite{GeBi},\cite{CG},\cite{GR} are the same as the ones we exhibit in
Appendix, except for the renormalization pointed out above. It should
however be added that only by virtue of such a change of normalization can
the locality property of the previous subsection be fulfilled.

\section{The $sl_n$ case}

It is easy to generalize the above results to the $sl_n$ case, at least
as far as the defining representation is concerned. For the Cartan
subalgebra we choose the following orthonormal basis
\ban
H_k\,=\, \frac {1} {\sqrt {k(k+1)}}\sum_{i=1}^{k}i(e_{i,i}-e_{i+1,i+1}),
\quad\quad\quad
k=1,...,n-1
\ean
The quantum $R$ matrix in defining representation of $sl_n$ is (see
\cite{JR})
\ban
R_{ij,kl}\,=\, \left\{ \brr{cc}
q^{\frac {n-1} {n}} \quad\quad &i=j=k=l \\
q^{-\frac {1}{n}}  \quad\quad & i=k,~ j=l,~i\neq j\\
q^{-\frac {1}{n}}x \quad\quad & i=l< j=k \\
0 \quad\quad&\mbox{\small\rm otherwise}\err\right.
\ean
where $x= q-q^{-1}$, as above. Let us denote by $A_i$ and $\bar
A_i,~i=1,...,n$ the diagonal elements $S_{ii}$ and $\bar S_{ii}$ of the
monodromy matrices $S$ and $\bar S$, respectively. The exchange algebra in
the Bloch wave basis is,

for $n>m$
\be
\brr{ll}
\psi^i_n \psi^i_m \,=\, q^{\frac {n-1} {n}} \psi_m^i \psi_n^i &
\\[12pt]
\psi^i_n \psi^j_m \,=\, q^{-\frac {1} {n}}\left[ \psi_m^j \psi_n^i +
x \frac {A_i} {A_j-A_i} \psi_m^i \psi_n^j\right], & i<j\label{exsln}
\\[12pt]
\psi^i_n \psi^j_m \,=\, q^{-\frac {1} {n}}\left[
\left(1-x^2 \frac {A_iA_j}{(A_i-A_j)^2}\right) \psi_m^j \psi_n^i +
x \frac {A_i} {A_i-A_j} \psi_m^i \psi_n^j\right], \quad\quad\quad & i>j
\err
\ee

for $n<m$

\be
\brr{ll}
\psi^i_n \psi^i_m \,=\, q^{-\frac {n-1} {n}} \psi_m^i \psi_n^i &
\\[12pt]
\psi^i_n \psi^j_m \,=\, q^{\frac {1} {n}}\left[ \psi_m^j \psi_n^i +
x \frac {A_j} {A_j-A_i} \psi_m^i \psi_n^j\right], & i<j\label{exsln2}
\\[12pt]
\psi^i_n \psi^j_m \,=\, q^{\frac {1} {n}}\left[
\left(1-x^2 \frac {A_iA_j}{(A_i-A_j)^2}\right) \psi_m^j \psi_n^i +
x \frac {A_j} {A_i-A_j} \psi_m^i \psi_n^j\right], \quad\quad\quad & i>j\,.
\err
\ee

Likewise we can write down the exchange algebra for $\bar \psi_n$ and construct
$\psi_n \bar \psi_n$. Periodicity of these objects is guaranteed in the
subspace
${\cal H}_0$ of the total Hilbert space ${\cal H}$ where the conditions
\be
A_i \bar A_i \,=\, q^{\frac {n+1}{n}} \label{perio}
\ee
are satisfied. As for locality, we find
\be
\relax [\psi_n \bar \psi_m, \psi_n\bar \psi_m]\,=\, x \sum_{i<j}
\frac{A_j\bar A_j- A_i\bar A_i}{(A_j-A_i)(\bar A_j-\bar A_I)}
\left( \psi_n^i\psi_m^j \bar \psi_n^j \bar \psi_m^i-
\psi_m^i\psi_n^j \bar \psi_m^j \bar \psi_n^i\right)\,.\label{locali}
\ee
So, also in this general case, the condition (\ref{perio}) guarantees
locality as well.

\newpage
\section*{Appendix}

\subsection*{A.1~The  $\psi$~$\psi$ exchange algebra}

In the defining representation we find (recall that $x=q-q^{-1}$)
\ban
\lefteqn{\mbox{{\em $\,\,$i)} ~~~case $n>m$}} \0\\[20pt]
&&\brr{l}
\psi_n^1\psi_m^1 \,=\, q^{{2\over 3}}\,\psi_m^1\psi_n^1
\err\0\\[12pt]
&&\brr{l}
\psi_n^1\psi_m^2 \,=\, q^{-{1\over 3}}\,\psi_m^2\psi_n^1 -
q^{-{1\over 3}}x{A\over B-A}\,\psi_m^1\psi_n^2
\err\0\\[12pt]
&&\brr{l}
\psi_n^1\psi_m^3 \,=\, q^{-{1\over 3}}\,\psi_m^3\psi_n^1 -
q^{-{1\over 3}}x{A\over C-A}\,\psi_m^1\psi_n^3
\err\0\\[12pt]
&&\brr{l}
\psi_n^2\psi_m^1 \,=\, q^{-{1\over 3}}\left[1-x ^2
{AB\over (B-A)^2}\right]\,\psi_m^1\psi_n^2 + q^{-{1\over 3}}x
{B\over B-A}\,\psi_m^2\psi_n^1
\err\0\\[12pt]
&&\brr{l}
\psi_n^2\psi_m^2 \,=\, q^{{2\over 3}}\,\psi_m^2\psi_n^2
\err\0\\[12pt]
&&\brr{l}
\psi_n^2\psi_m^3 \,=\, q^{-{1\over 3}}\,\psi_m^3\psi_n^2 -
q^{-{1\over 3}}x{B\over C-B}\,\psi_m^2\psi_n^3
\err\0\\[12pt]
&&\brr{l}
\psi_n^3\psi_m^1 \,=\, q^{-{1\over 3}}\left[1-x^2
{AC\over (C-A)^2}\right]\,\psi_m^1\psi_n^3 + q^{-{1\over 3}}x
{C\over C-A}\,\psi_m^3\psi_n^1
\err\0\\[12pt]
&&\brr{l}
\psi_n^3\psi_m^2 \,=\, q^{-{1\over 3}}\left[1-x^2
{BC\over (C-B)^2}\right]\,\psi_m^2\psi_n^3 + q^{-{1\over 3}}x
{C\over C-B}\,\psi_m^3\psi_n^2 \quad\quad\quad\quad
\err\0\\[12pt]
&&\brr{l}
\psi_n^3\psi_m^3 \,=\, q^{{2\over 3}}\,\psi_m^3\psi_n^3
\err\0\\[20pt]
\lefteqn{\mbox{{\em $\,$ii)} ~~~case $n=m$}} \\[20pt]
&&\brr{l}
\psi_n^1\psi_n^2 \,=\, {B-A \over qB-q^{-1}A }\,\psi_n^2\psi_n^1
\err\0\\[12pt]
&&\brr{l}
\psi_n^2\psi_n^3 \,=\, {C-B \over qC-q^{-1}B }\,\psi_n^3\psi_n^2
\err\0\\[12pt]
&&\brr{l}
\psi_n^1\psi_n^3 \,=\, {C-A \over qC-q^{-1}A }\,\psi_n^3\psi_n^1
\err\0\\[20pt]
\lefteqn{\mbox{{\em iii)} ~~~case $n<m$}} \\[20pt]
&&\brr{l}
\psi_n^1\psi_m^1 \,=\, q^{-{2\over 3}}\,\psi_m^1\psi_n^1
\err\0\\[12pt]
&&\brr{l}
\psi_n^1\psi_m^2 \,=\, q^{{1\over 3}}\,\psi_m^2\psi_n^1 -
q^{{1\over 3}}x{B\over B-A}\,\psi_m^1\psi_n^2
\err\0\\[12pt]
&&\brr{l}
\psi_n^1\psi_m^3 \,=\, q^{{1\over 3}}\,\psi_m^3\psi_n^1 -
q^{{1\over 3}}x{C\over C-A}\,\psi_m^1\psi_n^3
\err\0\\[12pt]
&&\brr{l}
\psi_n^2\psi_m^1 \,=\, q^{{1\over 3}}\left[1-x^2
{AB\over (B-A)^2}\right]\,\psi_m^1\psi_n^2 + q^{{1\over 3}}x
{A\over B-A}\,\psi_m^2\psi_n^1
\err\0\\[12pt]
&&\brr{l}
\psi_n^2\psi_m^2 \,=\, q^{-{2\over 3}}\,\psi_m^2\psi_n^2
\err\0\\[12pt]
&&\brr{l}
\psi_n^2\psi_m^3 \,=\, q^{{1\over 3}}\,\psi_m^3\psi_n^2 -
q^{{1\over 3}}x{C\over C-B}\,\psi_m^2\psi_n^3
\err\0\\[12pt]
&&\brr{l}
\psi_n^3\psi_m^1 \,=\, q^{{1\over 3}}\left[1-x^2
{AC\over (C-A)^2}\right]\,\psi_m^1\psi_n^3 + q^{{1\over 3}}x
{A\over C-A}\,\psi_m^3\psi_n^1
\err\0\\[12pt]
&&\brr{l}
\psi_n^3\psi_m^2 \,=\, q^{{1\over 3}}\left[1-x^2
{BC\over (C-B)^2}\right]\,\psi_m^2\psi_n^3 + q^{{1\over 3}}x
{B\over C-B}\,\psi_m^3\psi_n^2
\err\0\\[12pt]
&&\brr{l}
\psi_n^3\psi_m^3 \,=\, q^{-{2\over 3}}\,\psi_m^3\psi_n^3
\err
\ean
The exchange algebra $\psi^*~\psi^*$ can be obtained from this by simply
starring it, i.e. replacing everywhere $\psi, A,B,C$ by $\psi^*, A^*,B^*$ and
$C^*$, respectively.

The antichiral algebra $\bar \psi~\bar \psi$ can be obtained from the
above following the
recipe: in order to get the exchange relation of $\bar \psi_n^i~\bar \psi_m^j$,
take the exchange relation of $\psi_m^j~\psi_n^i$ and bar everything
including $A,B$ and $C$. For example, for $n>m$, we get
\ban
\bar \psi_n^1\bar \psi_m^2 \,=\, q^{{1\over 3}}\left[1-x^2
{{\bar A \bar B}\over (\bar B-\bar A)^2}\right]\,\bar \psi_m^2\bar\psi_n^1
+ q^{{1\over 3}}x
{\bar A\over {\bar B-\bar A}}\,\bar\psi_m^1\bar\psi_n^2
\ean

Of course the algebra $\bar \psi^*~\bar\psi^*$ is obtained by simply
``starring" the algebra $\bar \psi~\bar \psi$.

\subsection*{A.2~The  $\psi~\psi^*$ exchange algebra}

This algebra is given by
\ban
\lefteqn{\mbox{{\em $\,\,$i)} ~~~case $n>m$}} \\[20pt]
&&\brr{l}
\psi_n^1\psi_m^{*1} \,=\, q^{{1\over 3}}\,\psi_m^{*1}\psi_n^1
\err\0\\[12pt]
&&\brr{l}
\psi_n^1\psi_m^{*2} \,=\, q^{1\over 3}\,\psi_m^{*2}\psi_n^1
\err\0\\[12pt]
&&\brr{l}
\psi_n^1\psi_m^{*3} \,=\, q^{-{2\over 3}}\,\psi_m^{*3}\psi_n^1
-q^{-{2\over 3}}x{A\over B-A}\,\psi_m^{*2}\psi_n^{2} +
\\[10pt]
{}~~~~~~~~~~~~ + q^{-{2\over 3}}x
{A\over C-A}\,{qC-q^{-1}B\over C-B}\,\psi_m^{*1}\psi_n^3
\err\0\\[12pt]
&&\brr{l}
\psi_n^2\psi_m^{*1} \,=\, q^{{1\over 3}}\psi_m^{*1}\psi_n^2
\err\0\\[12pt]
&&\brr{l}
\psi_n^2\psi_m^{*2} \,=\, q^{-{2\over 3}}x{B\over B-A}
\,\psi_m^{*3}\psi_n^1 + q^{-{2\over 3}}\left[1-x^2{AB\over (B-A)^2}\right]
\,\psi_m^{*2}\psi_n^2 + \quad\quad\quad\quad
\\[10pt]
{}~~~~~~~~~~~~ - q^{-{2\over 3}}x{B\over C-B}\,{q^{-1}B-qA\over B-A}\,
       {qC-q^{-1}A\over C-A}\,\psi_m^{*1}\psi_n^3
\err\0\\[12pt]
&&\brr{l}
\psi_n^2\psi_m^{*3} \,=\, q^{1\over 3}\,\psi_m^{*3}\psi_n^2
\err\0\\[12pt]
&&\brr{l}
\psi_n^3\psi_m^{*1} \,=\, - q^{-{2\over 3}}x{C\over C-A}\,
{q^{-1}C-qB\over C-B}\,\psi_m^{*3}\psi_n^1 +
\\[10pt]
{}~~~~~~~~~~~~ + q^{-{2\over 3}}x{C\over C-B}\,
{qB-q^{-1}A\over B-A}\,{q^{-1}C-qA\over C-A}\,\psi_m^{*2}\psi_n^2 +
\\[10pt]
{}~~~~~~~~~~~~ + q^{-{2\over 3}}\left[1-x^2
{AC\over (C-A)^2}\right]\left[1-x^2{BC\over (C-B)^2}\right]\,
\psi_m^{*1}\psi_n^3
\err\0\\[12pt]
&&\brr{l}
\psi_n^3\psi_m^{*2} \,=\, q^{1\over 3}\,\psi_m^{*2}\psi_n^3
\err\0\\[12pt]
&&\brr{l}
\psi_n^3\psi_m^{*3} \,=\, q^{1\over 3}\,\psi_m^{*3}\psi_n^3
\err\0\\[20pt]
\lefteqn{\mbox{{\em iii)} ~~~case $n<m$}} \\[20pt]
&&\brr{l}
\psi_n^1\psi_m^{*1} \,=\, q^{-{1\over 3}}\,\psi_m^{*1}\psi_n^1
\err\0\\[12pt]
&&\brr{l}
\psi_n^1\psi_m^{*2} \,=\, q^{-{1\over 3}}\,\psi_m^{*2}\psi_n^1
\err\0\\[12pt]
&&\brr{l}
\psi_n^1\psi_m^{*3} \,=\, q^{{2\over 3}}\,\psi_m^{*3}\psi_n^1 -
q^{{2\over 3}}x{B\over B-A}\,\psi_m^{*2}\psi_n^2 +
\\[10pt]
{}~~~~~~~~~~~~ + q^{{2\over 3}}x{C\over C-A}\,{qC-q^{-1}B\over C-B}
\,\psi_m^{*1}\psi_n^3
\err\0\\[12pt]
&&\brr{l}
\psi_n^2\psi_m^{*1} \,=\, q^{-{1\over 3}}\,\psi_m^{*1}\psi_n^2
\err\0\\[12pt]
&&\brr{l}
\psi_n^2\psi_m^{*2} \,=\, q^{{2\over 3}}x{A\over B-A}\,
\psi_m^{*3}\psi_n^1 + q^{{2\over 3}}\left[1-x^2{AB\over(B-A)^2}\right]
\,\psi_m^{*2}\psi_n^2 +
\\[10pt]
{}~~~~~~~~~~~~ - q^{{2\over 3}}x{C\over C-B}\,{q^{-1}B-qA\over B-A}\,
{qC-q^{-1}A\over C-A}\,\psi_m^{*1}\psi_n^3
\err\0\\[12pt]
&&\brr{l}
\psi_n^2\psi_m^{*3} \,=\, q^{-{1\over 3}}\,\psi_m^{*3}\psi_n^2
\err\0\\[12pt]
&&\brr{l}
\psi_n^3\psi_m^{*1} \,=\,- q^{{2\over 3}}x{A\over C-A}\,
{q^{-1}C-qB\over C-B}\,\psi_m^{*3}\psi_n^1 +
\\[10pt]
{}~~~~~~~~~~~~ + q^{{2\over 3}}x{B\over C-B}\,{qB-q^{-1}A\over B-A}\,
{q^{-1}C-qA\over C-A}\,\psi_m^{*2}\psi_n^2 +
\\[10pt]
{}~~~~~~~~~~~~ + q^{{2\over 3}}\left[1-x^2
{AC\over (C-A)^2}\right]\left[1-x^2{CB\over (C-B)^2}\right]\,
\psi_m^{*1}\psi_n^3
\err\0\\[12pt]
&&\brr{l}
\psi_n^3\psi_m^{*2} \,=\, q^{-{1\over 3}}\,\psi_m^{*2}\psi_n^3
\err\0\\[12pt]
&&\brr{l}
\psi_n^3\psi_m^{*3} \,=\, q^{-{1\over 3}}\,\psi_m^{*3}\psi_n^3
\err
\ean
The exchange algebra for $n=m$ is the same as for $n>m$ with the RHS multiplied
by $q^{-\frac {1}{3}}$.

By ``starring" this algebra we obtain the algebra $\psi^*~\psi$.

The corresponding antichiral algebra $\bar \psi~\bar \psi^*$ can be obtained
according to the recipe: to get $\bar \psi_n^i~\bar \psi_m^{*j}$,
take $\psi_m^{*j}~\psi_n^i$ and bar everything including $A,B$ and $C$.

{}From this one can write down the algebra $\bar \psi^*~\bar \psi$.

\vskip1cm
{\bf Acknowledgements} One of us (L.B.) would like to thank the Instituto de
Fisica Teorica -- UNESP of S.Paulo for the kind hospitality extended to him
during the completion of this paper.

\end{document}